\documentclass[12pt]{iopart}
\usepackage{graphicx}
\usepackage{dcolumn}
\usepackage{bm}
\usepackage{hyperref}
\usepackage{cite}
\usepackage[utf8]{inputenc}
\usepackage[
top    = 2.75cm,
bottom = 3.5cm,
left   = 3.00cm,
right  = 2.50cm]{geometry}
\usepackage{fancyhdr}
\begin{document}

\title{Fabrication of hybrid molecular devices using multi-layer graphene break junctions} 

\author{J.O. Island\dag, A. Holovchenko\dag, M. Koole\dag, P.F.A. Alkemade\dag, M. Menelaou\ddag, N. Aliaga-Alcalde\S, E. Burzuri\dag,  H.S.J. van der Zant\dag}
\address{\dag Kavli Institute of Nanoscience, Delft University of Technology, Lorentzweg 1, 2628 CJ Delft, The Netherlands}
\address{\ddag Universitat de Barcelona, Facultat de Química, Diagonal 645, 08028 Barcelona (Spain)}
\address{\S Institució Catalana de Recerca I Estudis Avançats (ICREA) \& Institut de Ciència de Materials de Barcelona (ICMAB-CSIC), Campus de la UAB, 08193 Bellaterra (Spain)}

\ead{h.s.j.vanderzant@tudelft.nl}
\date{\today}

\begin{abstract}
We report on the fabrication of hybrid molecular devices employing multi-layer graphene (MLG) flakes which are patterned with a constriction using a helium ion microscope (HIM) or an oxygen plasma etch. The patterning step allows for the localization of a few-nanometer gap, created by electroburning, that can host single molecules or molecular ensembles. By controlling the width of the sculpted constriction, we regulate the critical power at which the electroburning process begins. We estimate the flake temperature given the critical power and find that at low powers it is possible to electroburn MLG with superconducting contacts in close proximity. Finally, we demonstrate the fabrication of hybrid devices with superconducting contacts and anthracene-functionalized copper curcuminoid molecules. This method is extendable to spintronic devices with ferromagnetic contacts and a first step towards molecular integrated circuits. 
\end{abstract}


\maketitle 
In recent years, molecular electronics has greatly benefited from the fabrication of nanometer spaced gold electrodes in the form of electromigrated and mechanical break junctions \cite{reed97, park99, kergueris99, park00, smit02, park02, tao06}. While these advances have led to a wealth of new physics in molecular devices, there are improvements to be made and a continuing focus in the field is toward making robust and reliable contact to single molecules\cite{Lortscher13}. One recent direction is the use of graphene or multi-layer graphene (MLG) as an electrode and contacting molecules by covalent bonding or $\pi-\pi$ stacking of aromatic rings \cite{bailey14, garcia13, wang11new, cao12}. This method resulted in observation of the gating of a single molecule device at room temperature \cite{prins11}. While gold atoms are mobile at room temperature \cite{Prins09, Strachan06}, MLG can be used as a stable intermediary material to contact molecules. In addition, two-dimensional MLG bridges better the size discrepancy between single molecules and bulky 3D electrodes where screening of the gate can reduce coupling to the molecule. Using MLG as an intermediary material also allows the option of contacting superconducting or ferromagnetic metals without changing the anchoring chemistry of the molecules. This opens the door to the fabrication of hybrid devices to study superconductivity and spintronics in single molecules. Up to now, the added complexity in fabricating hybrid superconducting-molecular devices has kept their exploration to a minimum \cite{winkelmann09}.

In this paper, we present further steps in the development of MLG electrodes with gold contacts and improve yield and stability through the localization of a constriction in an exfoliated flake. A predefined constriction using a helium ion microscope (HIM) to mill the flake or an oxygen plasma to etch the flake allows the flexibility to choose the location of a few-nm gap created by a process of electroburning. In comparison with as-exfoliated devices (without a constriction), MLG break junctions (with a constriction) show a 15\% increase in yield due to a reduction in abrupt breaks and reconnected junctions after electroburning \cite{burzuri12}. The key to using MLG as opposed to graphene or even few-layer graphene is two-fold: the conductance of MLG is independent of back-gate voltage making electron-transport measurements solely a study of the molecular characteristics of the device and not of the MLG electrodes, and we have found, as detailed in Ref. \cite{burzuri12}, that the highest yield for nanogap formation is in devices having a resistance of 500-1k$\Omega$ corresponding to devices with a MLG flake of $10$ nm thickness with a variation of approximately 5 nm. Furthermore, we explore the possibility of fabricating hybrid superconducting-molecular devices. The constriction results in a significant reduction in the power required to start the burning process making it possible to fabricate MLG break junctions with superconducting contacts. We use the superconducting alloy, molybdenum-rhenium (MoRe 60/40) which has a higher melting temperature ($T_M=1150$ $^{\circ}$C) as compared with commonly used superconductors such as aluminum ($T_M=660$ $^{\circ}$C) \cite{witcomb73}. Out of 26 fabricated MLG break junctions with MoRe contacts, 16 junctions resulted in tunnel gaps with an average low bias ($V_b = 10$ mV) resistance of $17\pm11$ G$\Omega$. 

A brief summary of the fabrication and electroburning process for MLG electrodes is given here which is detailed in Refs. \cite{prins11, burzuri12}. MLG devices are fabricated starting with heavily-doped Si wafers with a 285 nm-thick SiO$_{2}$ film. The Si substrate is used as a gate electrode for three-terminal devices. MLG flakes are deposited by mechanical exfoliation onto the  SiO$_{2}$ and flakes with a thickness of approximately $10$ nm are chosen by their color contrast under an optical microscope. Typical flakes have a deep purple color under white light illumination and a red channel contrast of about 16\% (see supplement information). Gold contacts, or MoRe contacts in the case of hybrid devices, are created by patterning with e-beam lithography and thin-film metal deposition. 

Electroburning is performed at room temperature and in air. A bias voltage up to $10$ V, corresponding to an electric field of more than $10^6$ V/m for typical devices, is ramped up across the flake. At large current densities, removal of carbon atoms in the MLG flake occurs due to the interaction with oxygen triggered by the high temperatures of Joule self heating \cite{prins11, burzuri12}. Feedback controlled software is used to control the voltage across the flake. When the conductance drops by 10\% the voltage is swept down at a rate of $0.1$ V/$\mu s$ to arrest the burning process and prevent the formation of a large gap. This cycle continues until the low bias ($V_b=10$ mV) resistance is greater than 10 G$\Omega$, independent of sample geometry. This resistance ensures the electroburning of any last remaining nanometer-sized graphene islands which have been shown to act as large addition energy quantum dots \cite{barreiro122}. By controlling the voltage in this way, a gap is burned across the entire flake with a width of a few nanometers at its closest point \cite{molina14}. 

\begin{figure}
\centerline{\includegraphics {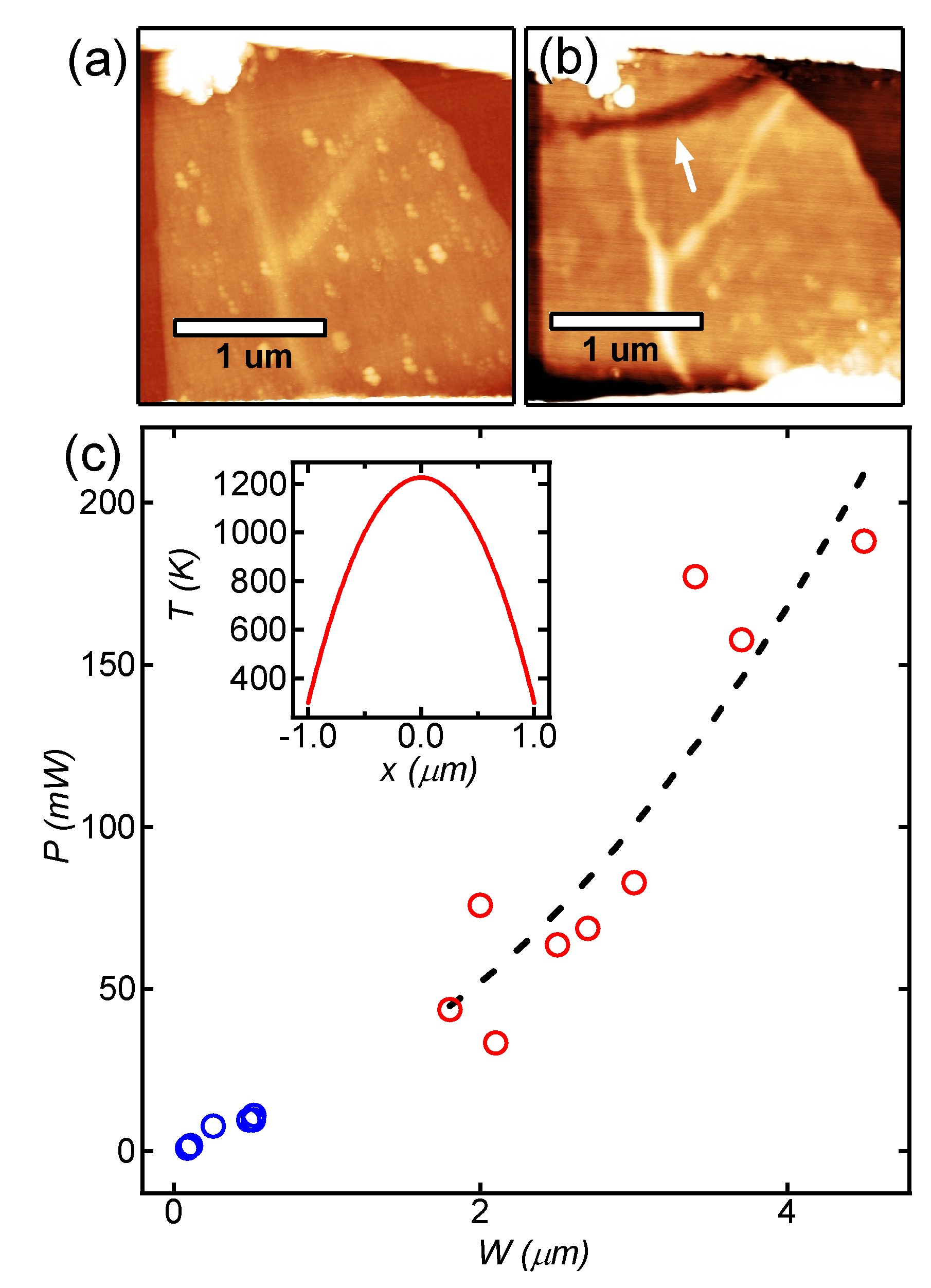}}
\caption{\label{} (a) AFM image of a MLG device before electroburning. The saturated white regions at the top and bottom are gold electrodes. (b) AFM image of the same MLG flake after electroburning. The gap is seen near the top electrode indicated by a white arrow. (c) Critical power required to start the electroburning vs. the width of the MLG flake. Red circles show the critical powers for as-exfoliated devices. Blue circles show the critical power for pre-patterned break junction devices using the HIM. The dashed line shows the calculated critical power as a function of width (see text). Inset shows the calculated temperature profile for the widest flake of $4.5$ $\mu$m.}
\end{figure}

In Fig. 1 we show the electroburning results of as-exfoliated MLG flakes. Fig. 1(a) shows an atomic force microscopy (AFM) image of a few-micron-wide MLG flake before electroburning. Without localization of a constriction, the electroburning process is believed to most frequently start along the edge of the flake at an unsaturated carbon atom site where a lower displacement energy ($7$ eV) is required to remove an atom from the lattice as compared to removing a saturated atom from the flake center ($30$ eV) \cite{barreiro12, warner09}. In Fig. 1(b) the resulting few-nm gap is shown near the top of the electrode (white arrow). Electroburning of as-exfoliated flakes often results in gaps with unpredictable locations. Here, the close proximity of the gap to the gold electrode could result in screening of an applied gate field. Ideal devices should have gaps located near the center of the flake where the gate coupling to the molecule is unaffected. 

In Fig. 1(c), the critical power required to start the electroburning process for the flake shown in Fig. 1(a) is plotted along with several other flakes (red circles). Critical powers reach hundreds of mW for the widest flakes and decrease with the flake width. Assuming comparable flake width (W) and length (L), the critical power as a function of width can be estimated by $P_c=I_c^2R=(j_cWt)^2R_s L/W\approx(j_ct)^2R_sW^2$, where we take $j_c=5.8\times10^7$ A/cm$^2$ as the average critical current density in these devices \cite{prins11} and $R_s=100$ $\Omega$/sq as the sheet resistance for MLG \cite{park12}. Using these values and an average flake thickness of $t=10$ nm, we plot this critical power relation as a dashed line in Fig. 1(c) where the scatter in the data is mostly due to differences in flake geometry and thickness. 

The electroburning is triggered by the high temperatures from Joule heating of the MLG flake. It is interesting to estimate the temperatures reached given the critical power of the electroburning process for as-exfoliated MLG devices. An estimate of the temperature distribution along the flake (in the direction of current) can be made using the 1D heat equation \cite{pop10, bae10, yigen13}:    

\begin{equation}\label{xx}
A\frac{\delta}{\delta x}\Bigg(k\frac{\delta T}{\delta x}\Bigg)+ p'_{x}-g(T-T_{0})=0, 
\end{equation}

\noindent where $A=Wt$ is the cross-sectional area of the flake, $k$ is the thermal conductivity of the MLG flake, $p'_{x}$ is the Joule heating rate in watts per unit length, and $g$ is the thermal conductance to the substrate per unit length and is calculated given the thermal resistance of the SiO$_{2}$ and Si substrate beneath the MLG flake ($g\approx1/[L(R_{ox}+R_{Si})]$, where $L$ is the length of the flake (in the direction of current), $R_{ox}=t_{ox}/(k_{ox}WL)$, where $k_{ox}=1.4$ WK$^{-1}$m$^{-1}$ and $R_{Si}=1/(2k_{si}(WL)^{1/2})$, where $k_{Si}=50$ WK$^{-1}$m$^{-1}$)\cite{bae10}. Heat convection to air is neglected as the heat convection coefficient for air ($10-100$ Wm$^{-2}$K$^{-1}$) is four orders of magnitude smaller than the coefficient for heat conduction to the substrate (calculated below). 

An analytic solution to the 1D heat equation is arrived at by assuming that $k$ is constant and that there is a uniform heat generation along the flake given by $p'_{x}\approx I^2R/L$. The temperature in the flake at position $x$ is then given by: 

\begin{equation}\label{heat}
T(x)=T_0+\frac{p'_{x}}{g}\Bigg(1-\frac{cosh(x/L_H)}{cosh(L/2L_H)}\Bigg).
\end{equation}

\noindent Here, $T_0=300$ $K$, is the temperature of the electrodes and $L_H=(kA/g)^{1/2}$ is the characteristic healing length along the flake and is a measure of lateral heat dissipation in the flake. Using Eqn. \ref{heat}, we estimate the critical temperature distribution for the widest as-exfoliated flake in Fig. 1(c) ($W=4.5$ $\mu$m, $L=2$ $\mu$m) with a critical power of nearly $190$ mW. We take the thermal conductivity of MLG ($k=1000$ $WK^{-1}m^{-1}$) which is found for flakes of 8 layers \cite{ghosh10}. From the length and width of the flake, we calculate the thermal resistance of the Si substrate $R_{Si}\approx3\times10^3$ K/W, the thermal resistance of the oxide layer $R_{ox}\approx2\times10^4$ K/W, the thermal conductance per unit length $g\approx19$ WK$^{-1}$m$^{-1}$, and the characteristic healing length for this device, $L_H=(kA/g)^{1/2}=1.5$ $\mu$m. The healing length is larger than estimates for graphene ($L_H\approx 0.2$ $\mu$m on SiO$_{2}$) due to the thicker cross sectional area\cite{bae10}. The temperature distribution is plotted in the inset of Fig. 1(c). The center of the flake is expected to reach temperatures of $\approx1200$ K. Typical device dimensions are larger but comparable to the healing length making the 1D diffusion equation a sufficient estimate of the temperature distribution in the flake\cite{bae10}. A more accurate calculation can be made using finite element methods with COMSOL MULTIPHYSICS v4.3b to account for the small lateral heat dissipation (see supplemental information). The high flake temperatures prove detrimental to the fabrication of MLG electrodes with MoRe contacts, demonstrated below. These temperatures can be reduced by pre-patterning the flake to make MLG break junctions with decreased widths.

\begin{figure}
\centerline{\includegraphics[width=4in]{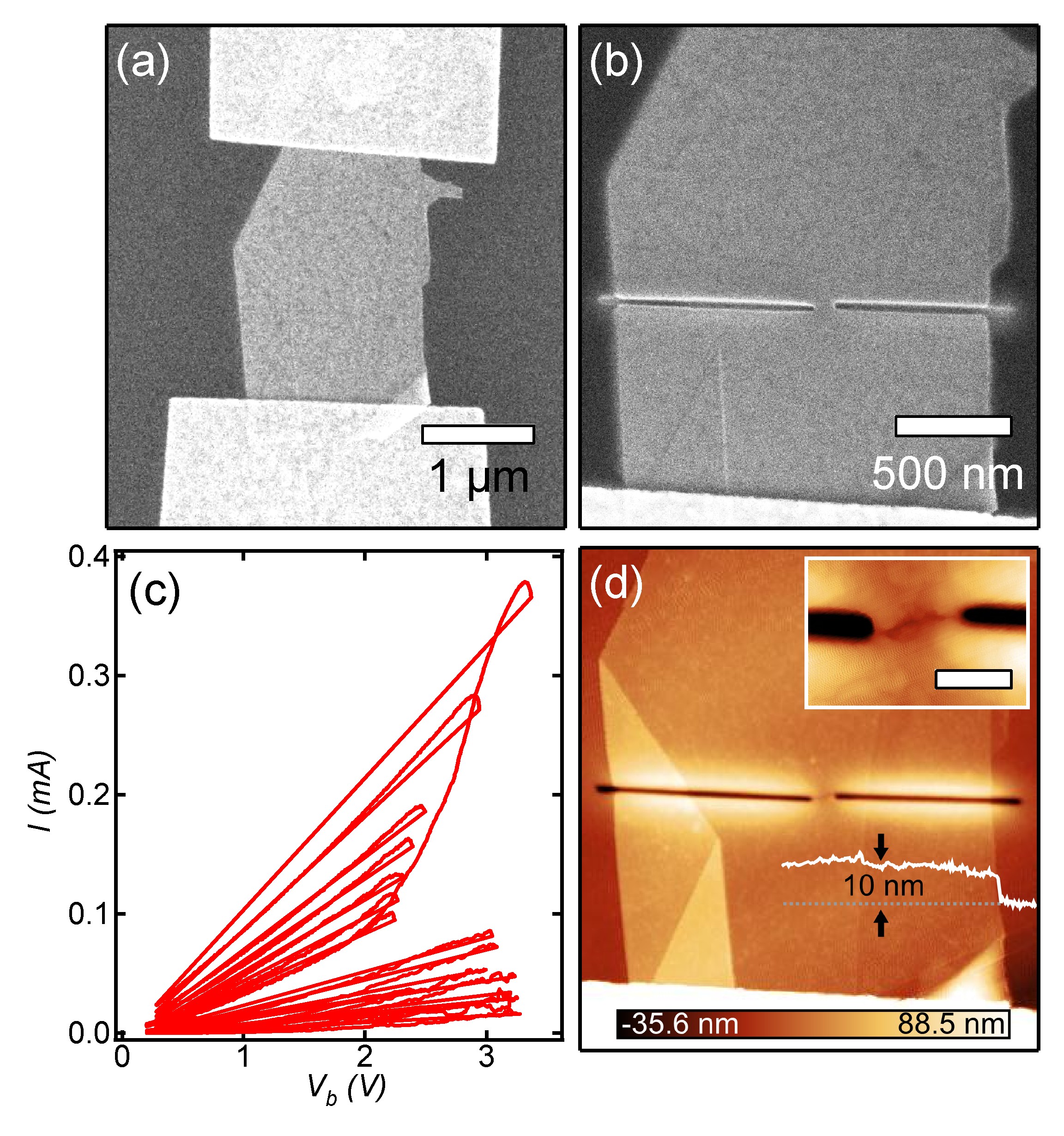}}
\caption{\label{} (a) Helium ion microscopy (HIM) image of a MLG device before milling between two Au electrodes. (b) HIM image of the same device after milling. The resulting constriction is $100$ nm wide. (c) $I-V_b$ characteristics for the electroburning of the constriction in (b). (d) AFM image of the same device after electroburning. The inset shows the final gap across the constriction. The scale bar is $100$ nm.}
\end{figure}

The electroburning of MLG break junctions is first demonstrated with gold contacts in Fig. 2.  Patterning of the MLG devices before electroburning is achieved using a HIM or by oxygen plasma etch. For oxygen plasma etching, we use a pressure of $50$ $\mu$Bar with a flow of 25 sccm O$_2$ at $20$ W power for 30 secs. The HIM is well suited for high resolution nanofabrication and unlike oxygen plasma etching, HIM patterning does not require an e-beam lithography step which makes the fabrication process cleaner and faster \cite{alkemade12}. In comparison with gallium ion beam milling, HIM patterning allows higher resolution and results in less damage to the Si/SiO$_{2}$ substrate \cite{bell09}. Patterning of graphene using a HIM has been achieved in several forms \cite{boden11, lemme09, zhou10}. Here, we use the HIM to position the location of the final few nm gap and reduce the critical power required to burn the gap. In order to minimize particle contamination of the chip during the milling process, the sample chamber is cleaned overnight with an integrated oxygen plasma cleaner. Imaging of the flake is made with lower magnification in order to prevent damage before milling. 

Figure 2(a) is a HIM image of a MLG flake before milling. This image is used to pattern rectangles that define the areas where the flake will be milled. After defining the milling area, the beam is focused on the edge of the flake and moved to the pattern. For ion beam milling an acceleration voltage of $25$ kV is used with a beam current of $1$ pA. Typical line doses are about $120$ nC/cm. Two rectangular patterns, one on each side of the flake, become $10$ nm-wide trenches after direct helium ion beam writing forming the final constriction in the center of the flake. In Fig. 2(b) we show the device after milling with the HIM. The resulting constriction is about $100$ nm wide. The trench width and depth ($\approx40$ nm from AFM scan) ensure no residual connection between the milled multi-layer graphene edges. While the trench does extend into the oxide, we find no evidence for substrate damage or gate leakage after milling. 

After pre-patterning with the HIM, electroburning is performed as before but the constriction results in a significant reduction of the critical power required to start the process. Fig. 2(c) shows the $I-V_b$ characteristics of the electroburning process for the patterned device in Fig. 2(b). The critical power is $\approx1$ mW which is more than an order of magnitude smaller than the narrowest as-exfoliated MLG device in Fig. 1(c). Critical powers for devices with wider pre-patterned constrictions are also shown in Fig. 1(c) (blue circles). Critical powers steadily increase to $11$ mW for constriction widths up to $500$ nm. In Fig. 2(d) we show an AFM image of the device after electroburning. The inset shows the gap formed across the constriction. 

\begin{figure}
\centerline{\includegraphics{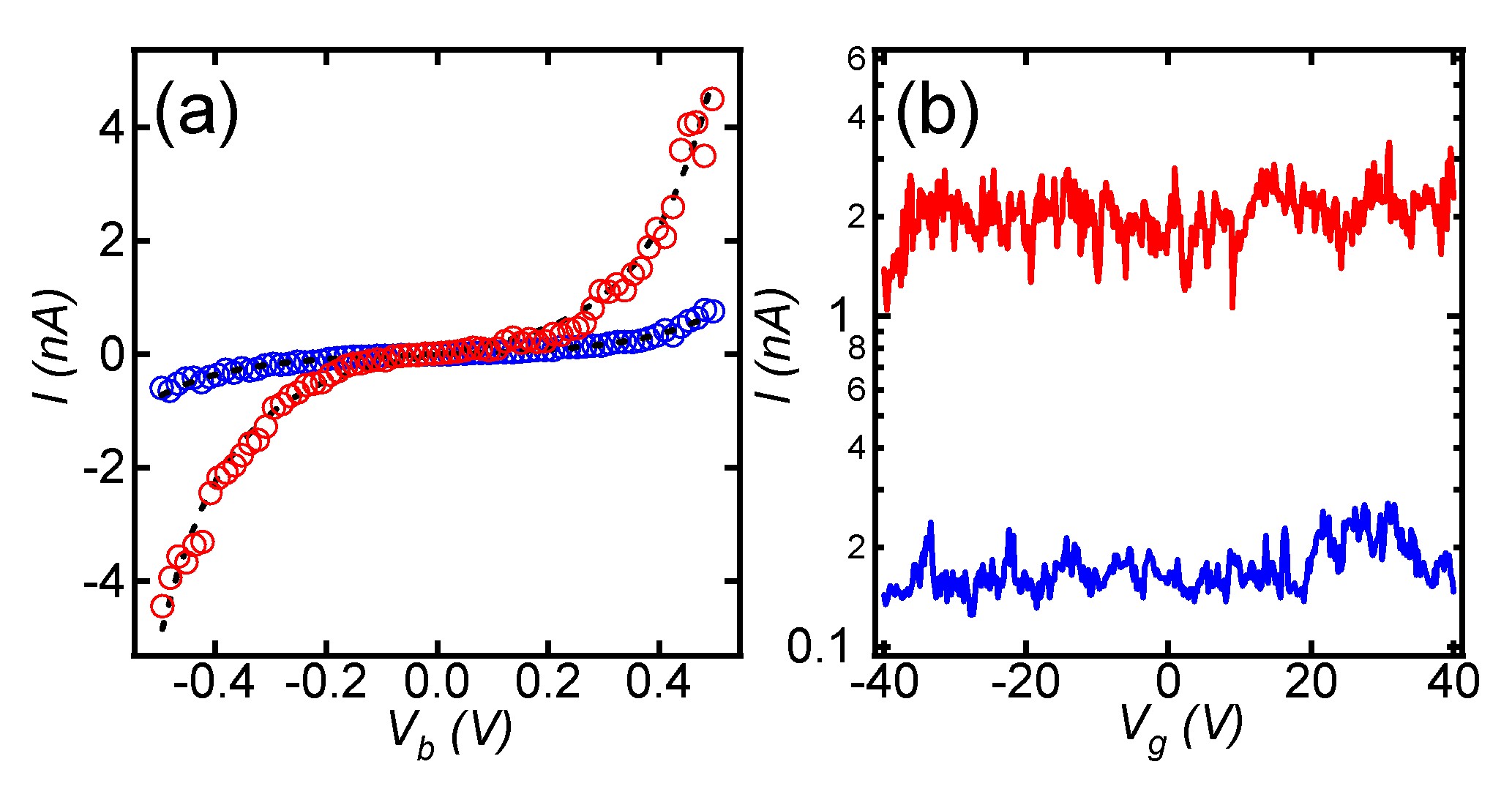}}
\caption{\label{} Electrical characterization after electroburning for two typical devices. (a) A characteristic tunneling curve is measured for the device after electroburning (red circles for device 1, blue circles for device 2). The Simmons model fit is plotted along with the tunnel current (black dashed line). (b) $I-V_g$ characteristics for the devices after electroburning at a bias voltage of $500$ mV.}
\end{figure}

We perform electrical measurements at room temperature and in vacuum to characterize the tunnel gap of the MLG break junctions. In Fig. 3 we show the $I-V_b$ and $I-V_g$ characteristics for two typical devices after electroburning. A tunneling current is measured through the vacuum barrier and shown in Fig. 3(a) (red and blue circles for devices 1 and 2, respectively). From the tunneling current, we estimate the gap distance using the Simmons model for tunneling through a thin insulating film\cite {simmons63, prins11}. 

\begin{equation}\label{heatequa}
I_{A, \phi, d}(V) =A\frac{e}{2\pi h d^2}\Bigg((\phi - \mu_L)e^{-2d\frac{\sqrt{2m(\phi-\mu_L)}}{\hbar}}-(\phi - \mu_R)e^{-2d\frac{\sqrt{2m(\phi-\mu_R)}}{\hbar}}\Bigg)
\end{equation}

\noindent Here, $\phi$ is the potential barrier height, $A$ is the cross sectional tunneling area, $d$ is the gap distance, and $\mu_L=+eV/2$ ($\mu_R=-eV/2$) are the chemical potentials in the left (right) lead. The fit using the Simmons model is plotted (black dashed line) along with the data in Fig. 3(a) where we have set $A=1000$ nm$^2$ given the cross sectional area of the fabricated constriction ($W=100$ nm, $t=10$ nm). From the fit, we extract a tunnel gap distance of $d\approx 2.1$ nm and a barrier height of $\phi\approx0.52$ eV for device 1 ($d\approx 2.1$ nm and $\phi\approx0.65$ eV for device 2). The low barrier height, as compared with that of graphene ($4.5$ eV), is common for the Simmons model when applied to small junctions (e.g. Au nanogaps) but could also include a barrier lowering due to water or OH groups absorbed at the gap edge \cite{mangin09, curtis12, wang11}. While the barrier height has been found to deviate from the expected value, scanning tunneling microscopy studies of Au nanogaps have shown a consistency between the extracted gap size from the Simmons fit and the actual electrode separation \cite{steinmann04}. We expect that this is the case for MLG nanogaps as well but further SEM studies should be carried out to verify this. In Fig. 3(b) we plot the $I-V_g$ characteristics for the empty tunnel gap at a bias voltage of $500$ mV. The current is independent of back-gate voltage.

\begin{figure}
\centerline{\includegraphics[width=3.35in]{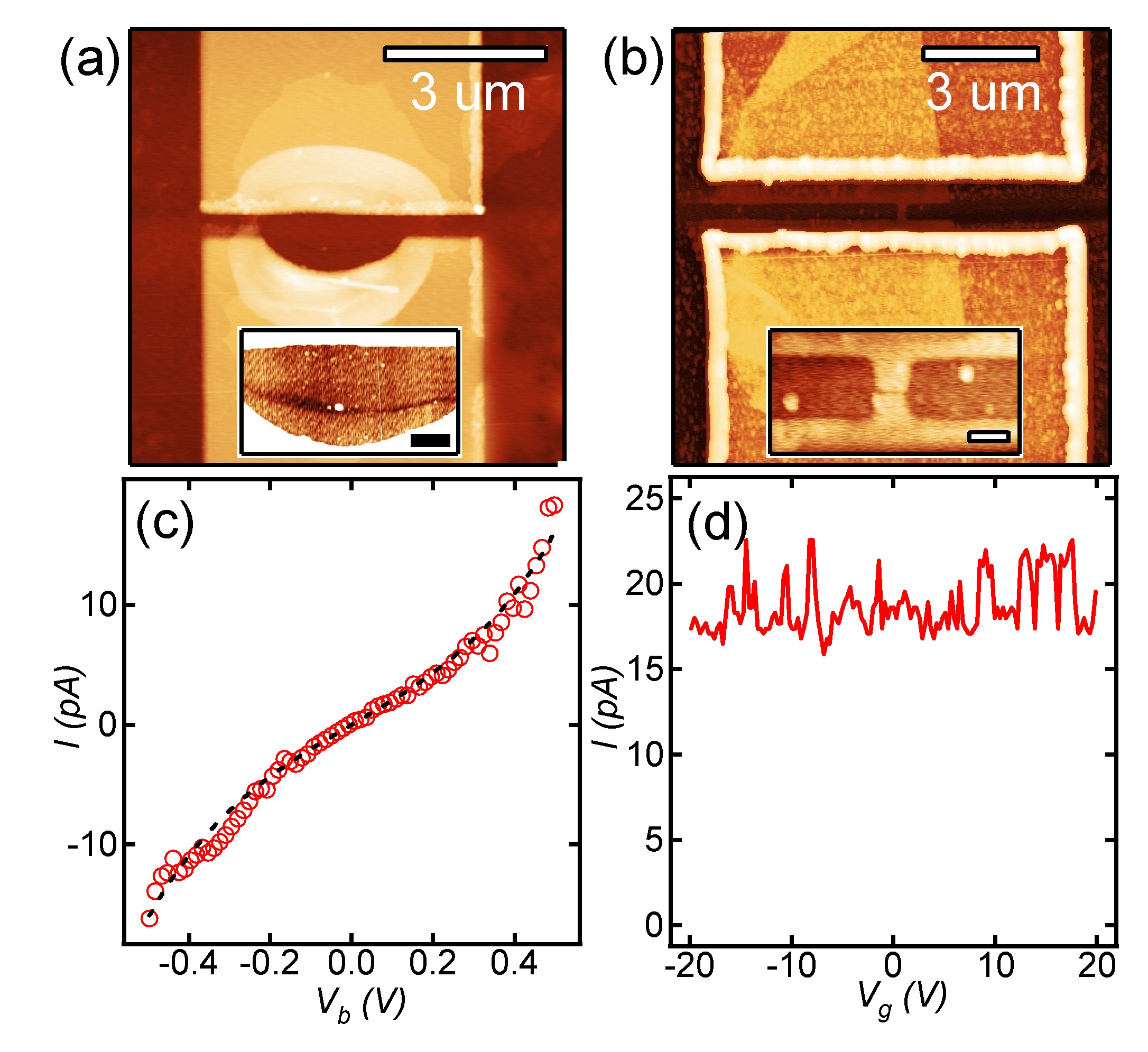}}
\caption{\label{} (a) AFM image of a device without prepatterning. The high temperature of the electroburning process deforms the lower electrode. Inset shows the location of the nanogap near the deformaded electrode (scale bar $500$ nm). (b) AFM image of an electroburned MLG break junction pre-patterned using an oxygen plasma etch. The inset shows the resulting constriction from etching (scale bar $100$ nm). The width is $\approx 80$ nm. (c) $I-V_b$ characteristics of the device in (b) (red circles) as well as a fit of the Simmons model (black dashed line). (d) $I-V_g$ characteristics of the device in (b) at a bias voltage of $500$ mV.}
\end{figure}

The high temperatures from electroburning as-exfoliated flakes are detrimental to the fabrication of MLG break junctions with superconducting contacts. Moreover, to achieve proximity effect in MLG for hybrid devices, channel lengths of $<0.5$ $\mu$m are needed \cite {kanda10}. The combination of high flake temperature and short channel length result in melting of the MoRe alloy for the wider ($>3$ $\mu m$) as-exfoliated devices. This can be seen in Fig. 4(a) which is an AFM image of an as-exfoliated flake device after electroburning. Deformation of the lower contact can be seen which is near the gap shown in the inset of Fig. 4(a). By pre-patterning the MLG flake, the electroburning process can be started at much lower critical powers. In Fig. 4(b) we show an example of an electroburned MLG break junction patterned by oxygen plasma etch with MoRe contacts. The MoRe contacts are unaffected and the gap can be seen located at the defined constriction in the inset of Fig. 4(b). In Figs. 4(c) and 4(d) we show the $I-V_b$, $I-V_g$ characteristics for the MLG electrodes with MoRe contacts shown in Fig. 4(b). Again, the Simmons model is fit to the tunnel current setting $A=800$ nm$^2$ given the cross sectional area of the constriction ($W=80$ nm, $t=10$ nm) and extract a gap distance of $d\approx1.6$ nm, a barrier height of $\phi\approx1.4$ eV. For the 16 devices that resulted in tunnel gaps after electroburning, the average tunnel gap from the Simmons fit is $2.1\pm0.3$ nm and the average barrier height is $0.89\pm0.3$ eV. 

To demonstrate the fabrication of hybrid devices, we have performed deposition of anthracene-functionalized copper curcuminoid molecules ([(Phen)CuCl(9Accm)]) on the MoRe MLG break junctions \cite{aliaga10}. Fig. 5(a) shows a scheme of the molecule. 
\begin{figure}
\centerline{\includegraphics[]{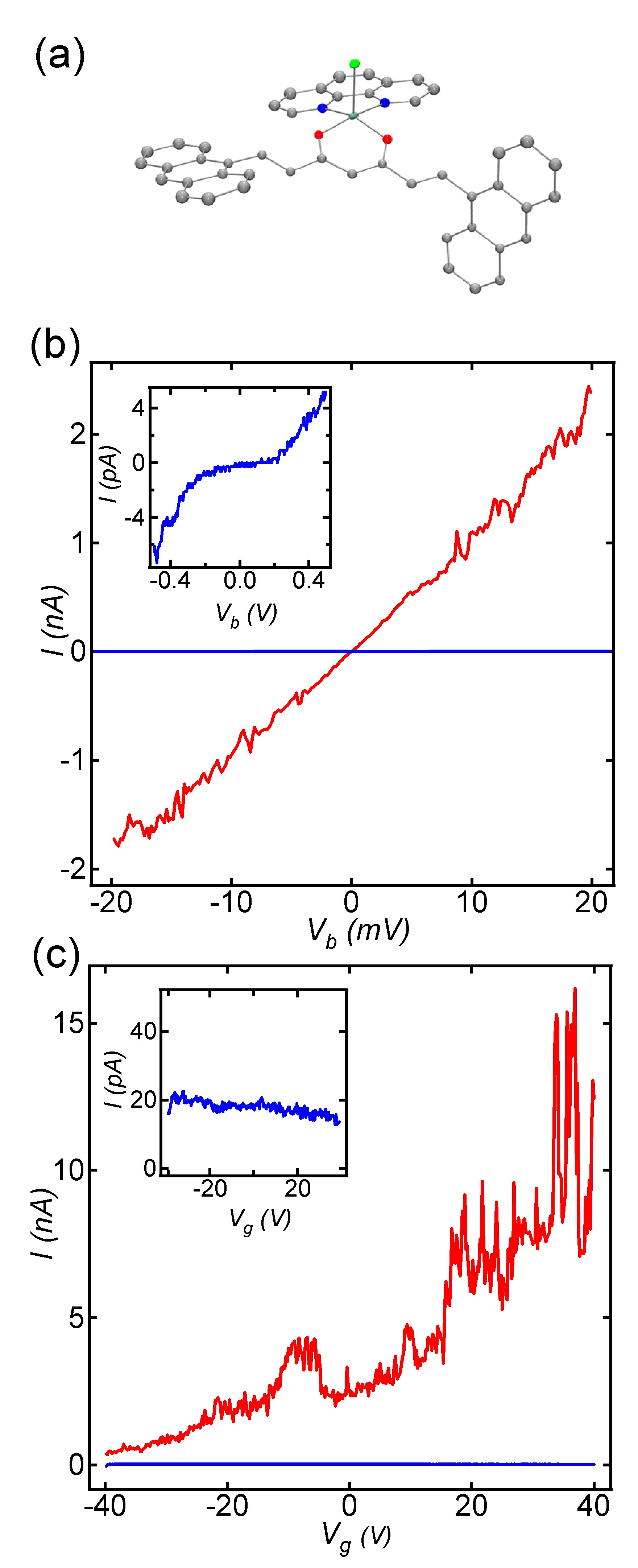}}
\caption{\label{} MoRe MLG break junction before and after molecule deposition. (a) Ortep representation of the copper curcuminoid molecule [(Phen)CuCl(9Accm)]. Phen stands for phenantroline and 9Accm corresponds to the curcuminoid ligand that contains anthracene groups. In the picture the copper atom is in light blue, the carbon atoms are grey, the oxygen atoms are red, the N atoms are blue and the Cl atom is green. Hydrogen atoms are not shown for simplification.  (b) $I-V_b$ characteristics before (inset) and after molecule deposition. (c) $I-V_g$ characteristics before (inset) and after molecule deposition. The $I-V_b$ and $I-V_g$ characteristics before deposition are also plotted in the main panel of (b) and (c) (blue lines) for comparison.}
\end{figure}
\noindent Contact to the molecule is mediated through the interaction of two $\pi$-systems; the $\pi$-conjugated system of the anthracene groups on the molecule and that of the MLG break junction. For deposition, a powder of synthesized molecules is dissolved in dichloromethane (CH$_{2}$Cl$_{2}$) at a concentration of 0.1 mM. The electroburned junctions are kept in the solution for 30 mins which results in a sub-monolayer of molecules on the surface of the Si/SiO$_{2}$ substrate \cite{prins11}. Of the remaining sixteen tunnel junctions after electroburning, one junction showed a significant change in the $I-V_b$ and $I-V_g$ characteristics after deposition. The inset of Fig. 5(b) shows the tunneling characteristics of the junction (low bias resistance of 14 G$\Omega$) and the inset of Fig. 5(c) shows that the current is virtually independent of gate voltage before deposition. After deposition, the $I-V_b$ characteristics become linear and the low bias resistance is 9 M$\Omega$. In addition, the current shows a strong dependence on gate voltage, changing by two orders of magnitude over 80 Volts on the back-gate. To ascertain the presence of a single molecule or a few molecules in the junction, further transport measurements can be performed at low temperature.

In conclusion, we have studied the yield and stability of MLG break junctions for use in hybrid molecular devices. Using HIM milling or an oxygen plasma etch to localize a constriction in the MLG flake, we control the position of a few nm gap created by electroburning. The critical power needed to start the electroburning process can be controlled by the width of the constriction. For lower powers ($\approx1$ mW), we fabricate MLG break junctions with MoRe contacts. Using these hybrid devices, we aim to study superconducting effects in single molecule transport. 

The authors would like to thank Dr. Emile van Veldhoven (TNO) for his help with the HIM and Dr. Andrés Castellanos-Gómez for helpful discussions and manuscript preparation. N.A.-A. would like to thank the Spanish MCI (CTQ2012-32247). This work was supported by The Dutch organization for Fundamental Research on Matter (FOM), NWO/OCW, the Dutch Technology Foundation STW, and by the EU FP7 project 618082 ACMOL and advanced ERC grant (Mols@Mols).

\bibliography{bib}

\begin{thebibliography}{10}

\bibitem{reed97}
Mark~A Reed, C~Zhou, CJ~Muller, TP~Burgin, and JM~Tour.
\newblock Conductance of a molecular junction.
\newblock {\em Science}, 278(5336):252--254, 1997.

\bibitem{park99}
Hongkun Park, Andrew~KL Lim, A~Paul Alivisatos, Jiwoong Park, and Paul~L
  McEuen.
\newblock Fabrication of metallic electrodes with nanometer separation by
  electromigration.
\newblock {\em Applied Physics Letters}, 75(2):301--303, 1999.

\bibitem{kergueris99}
C~Kergueris, J-P Bourgoin, S~Palacin, D~Esteve, C~Urbina, M~Magoga, and
  C~Joachim.
\newblock Electron transport through a metal-molecule-metal junction.
\newblock {\em Physical Review B}, 59(19):12505, 1999.

\bibitem{park00}
Hongkun Park, Jiwoong Park, Andrew~KL Lim, Erik~H Anderson, A~Paul Alivisatos,
  and Paul~L McEuen.
\newblock Nanomechanical oscillations in a single-c60 transistor.
\newblock {\em Nature}, 407(6800):57--60, 2000.

\bibitem{smit02}
RHM Smit, Y~Noat, C~Untiedt, ND~Lang, MC~Van~Hemert, and JM~Van~Ruitenbeek.
\newblock Measurement of the conductance of a hydrogen molecule.
\newblock {\em Nature}, 419(6910):906--909, 2002.

\bibitem{park02}
Jiwoong Park, Abhay~N Pasupathy, Jonas~I Goldsmith, Connie Chang, Yuval Yaish,
  Jason~R Petta, Marie Rinkoski, James~P Sethna, H{\'e}ctor~D Abru{\~n}a,
  Paul~L McEuen, et~al.
\newblock Coulomb blockade and the kondo effect in single-atom transistors.
\newblock {\em Nature}, 417(6890):722--725, 2002.

\bibitem{tao06}
NJ~Tao.
\newblock Electron transport in molecular junctions.
\newblock {\em Nature Nanotechnology}, 1(3):173--181, 2006.

\bibitem{Lortscher13}
Emanuel L{\"o}rtscher.
\newblock Wiring molecules into circuits.
\newblock {\em Nature nanotechnology}, 8(6):381--384, 2013.

\bibitem{bailey14}
Steven Bailey, David Visontai, Colin~J. Lambert, Martin~R. Bryce, Harry
  Frampton, and David Chappell.
\newblock A study of planar anchor groups for graphene-based single-molecule
  electronics.
\newblock {\em The Journal of Chemical Physics}, 140(5), 2014.

\bibitem{garcia13}
V{\'\i}ctor~Manuel Garc{\'\i}a-Su{\'a}rez, R~Ferrad{\'a}s, D~Carrascal, and
  Jaime Ferrer.
\newblock Universality in the low-voltage transport response of molecular wires
  physisorbed onto graphene electrodes.
\newblock {\em Physical Review B}, 87(23):235425, 2013.

\bibitem{wang11new}
Gunuk Wang, Yonghun Kim, Minhyeok Choe, Tae-Wook Kim, and Takhee Lee.
\newblock A new approach for molecular electronic junctions with a multilayer
  graphene electrode.
\newblock {\em Advanced Materials}, 23(6):755--760, 2011.

\bibitem{cao12}
Yang Cao, Shaohua Dong, Song Liu, Li~He, Lin Gan, Xiaoming Yu, Michael~L
  Steigerwald, Xiaosong Wu, Zhongfan Liu, and Xuefeng Guo.
\newblock Building high-throughput molecular junctions using indented graphene
  point contacts.
\newblock {\em Angewandte Chemie}, 124(49):12394--12398, 2012.

\bibitem{prins11}
Ferry Prins, Amelia Barreiro, Justus~W. Ruitenberg, Johannes~S. Seldenthuis,
  Núria Aliaga-Alcalde, Lieven M.~K. Vandersypen, and Herre S.~J. van~der
  Zant.
\newblock Room-temperature gating of molecular junctions using few-layer
  graphene nanogap electrodes.
\newblock {\em Nano Letters}, 11(11):4607--4611, 2011.

\bibitem{Prins09}
F.~Prins, T.~Hayashi, B.~J.~A. de~Vos~van Steenwijk, B.~Gao, E.~A. Osorio,
  K.~Muraki, and H.~S.~J. van~der Zant.
\newblock Room-temperature stability of pt nanogaps formed by self-breaking.
\newblock {\em Applied Physics Letters}, 94(12):123108, 2009.

\bibitem{Strachan06}
Douglas~R. Strachan, Deirdre~E. Smith, Michael~D. Fischbein, Danvers~E.
  Johnston, Beth~S. Guiton, Marija Drndić, Dawn~A. Bonnell, and Alan~T.
  Johnson.
\newblock Clean electromigrated nanogaps imaged by transmission electron
  microscopy.
\newblock {\em Nano Letters}, 6(3):441--444, 2006.

\bibitem{winkelmann09}
Clemens~B Winkelmann, Nicolas Roch, Wolfgang Wernsdorfer, Vincent Bouchiat, and
  Franck Balestro.
\newblock Superconductivity in a single-c60 transistor.
\newblock {\em Nature Physics}, 5(12):876--879, 2009.

\bibitem{burzuri12}
Enrique Burzur{\'\i}, Ferry Prins, and Herre~SJ van~der Zant.
\newblock Characterization of nanometer-spaced few-layer graphene electrodes.
\newblock {\em Graphene}, 1:26, 2012.

\bibitem{witcomb73}
MJ~Witcomb and D~Dew-Hughes.
\newblock The $\sigma$-phase in molybdenum-rhenium; microstructure and
  superconducting hysteresis.
\newblock {\em Journal of the Less Common Metals}, 31(2):197--209, 1973.

\bibitem{barreiro122}
Amelia Barreiro, Herre~SJ van~der Zant, and Lieven~MK Vandersypen.
\newblock Quantum dots at room temperature carved out from few-layer graphene.
\newblock {\em Nano letters}, 12(12):6096--6100, 2012.

\bibitem{molina14}
Aday~J Molina-Mendoza, Jos{\'e}~G Rodrigo, Joshua Island, Enrique Burzuri,
  Gabino Rubio-Bollinger, Herre~SJ van~der Zant, and Nicol{\'a}s Agra{\"\i}t.
\newblock Note: Long-range scanning tunneling microscope for the study of
  nanostructures on insulating substrates.
\newblock {\em Review of Scientific Instruments}, 85(2):026105, 2014.

\bibitem{barreiro12}
A.~Barreiro, F.~Börrnert, M.~H. Rümmeli, B.~Büchner, and L.~M.~K.
  Vandersypen.
\newblock Graphene at high bias: Cracking, layer by layer sublimation, and
  fusing.
\newblock {\em Nano Letters}, 12(4):1873--1878, 2012.

\bibitem{warner09}
Jamie~H Warner, Mark~H R{\"u}mmeli, Ling Ge, Thomas Gemming, Barbara Montanari,
  Nicholas~M Harrison, Bernd B{\"u}chner, and G~Andrew~D Briggs.
\newblock Structural transformations in graphene studied with high spatial and
  temporal resolution.
\newblock {\em Nature nanotechnology}, 4(8):500--504, 2009.

\bibitem{park12}
Jang-Ung Park, SungWoo Nam, Mi-Sun Lee, and Charles~M Lieber.
\newblock Synthesis of monolithic graphene--graphite integrated electronics.
\newblock {\em Nature materials}, 11(2):120--125, 2012.

\bibitem{pop10}
Eric Pop.
\newblock Energy dissipation and transport in nanoscale devices.
\newblock {\em Nano Research}, 3(3):147--169, 2010.

\bibitem{bae10}
Myung-Ho Bae, Zhun-Yong Ong, David Estrada, and Eric Pop.
\newblock Imaging, simulation, and electrostatic control of power dissipation
  in graphene devices.
\newblock {\em Nano letters}, 10(12):4787--4793, 2010.

\bibitem{yigen13}
S.~Yi\ifmmode~\breve{g}\else \u{g}\fi{}en, V.~Tayari, J.~O. Island, J.~M.
  Porter, and A.~R. Champagne.
\newblock Electronic thermal conductivity measurements in intrinsic graphene.
\newblock {\em Phys. Rev. B}, 87:241411, 2013.

\bibitem{ghosh10}
Suchismita Ghosh, Wenzhong Bao, Denis~L Nika, Samia Subrina, Evghenii~P
  Pokatilov, Chun~Ning Lau, and Alexander~A Balandin.
\newblock Dimensional crossover of thermal transport in few-layer graphene.
\newblock {\em Nature Materials}, 9(7):555--558, 2010.

\bibitem{alkemade12}
Paul~FA Alkemade, Emma~M Koster, Emile van Veldhoven, and Diederik~J Maas.
\newblock Imaging and nanofabrication with the helium ion microscope of the van
  leeuwenhoek laboratory in delft.
\newblock {\em Scanning}, 34(2):90--100, 2012.

\bibitem{bell09}
D~C Bell, M~C Lemme, L~A Stern, J~R Williams, and C~M Marcus.
\newblock Precision cutting and patterning of graphene with helium ions.
\newblock {\em Nanotechnology}, 20(45):455301, 2009.

\bibitem{boden11}
{S.A.} Boden, Z.~Moktadir, {D.M.} Bagnall, H.~Mizuta, and {H.N.} Rutt.
\newblock Focused helium ion beam milling and deposition.
\newblock {\em Microelectronic Engineering}, 88(8):2452--2455, August 2011.

\bibitem{lemme09}
Max~C. Lemme, David~C. Bell, James~R. Williams, Lewis~A. Stern, Britton W.~H.
  Baugher, Pablo Jarillo-Herrero, and Charles~M. Marcus.
\newblock Etching of graphene devices with a helium ion beam.
\newblock {\em {ACS} Nano}, 3(9):2674--2676, 2009.

\bibitem{zhou10}
Yong Zhou and Kian~Ping Loh.
\newblock Making patterns on graphene.
\newblock {\em Advanced Materials}, 22(32):3615--3620, 2010.

\bibitem{simmons63}
John~G Simmons.
\newblock Generalized formula for the electric tunnel effect between similar
  electrodes separated by a thin insulating film.
\newblock {\em Journal of Applied Physics}, 34:1793, 1963.

\bibitem{mangin09}
A~Mangin, A~Anthore, ML~Della~Rocca, E~Boulat, and P~Lafarge.
\newblock Reduced work functions in gold electromigrated nanogaps.
\newblock {\em Physical Review B}, 80(23):235432, 2009.

\bibitem{curtis12}
Kellye~S Curtis, Christopher~JB Ford, David Anderson, Harvey~E Beere, Ian
  Farrer, David~A Ritchie, and Geraint~AC Jones.
\newblock Reduced tunnel-barrier height in sub-10 nm au nanoelectrodes.
\newblock In {\em Nanotechnology (IEEE-NANO), 2012 12th IEEE Conference on},
  pages 1--9. IEEE, 2012.

\bibitem{wang11}
Weiliang Wang and Zhibing Li.
\newblock Potential barrier of graphene edges.
\newblock {\em Journal of Applied Physics}, 109(11):114308, 2011.

\bibitem{steinmann04}
Philipp Steinmann and JMR Weaver.
\newblock Fabrication of sub-5nm gaps between metallic electrodes using
  conventional lithographic techniques.
\newblock {\em Journal of Vacuum Science \& Technology B}, 22(6):3178--3181,
  2004.

\bibitem{kanda10}
A~Kanda, T~Sato, H~Goto, H~Tomori, S~Takana, Y~Ootuka, and K~Tsukagoshi.
\newblock Dependence of proximity-induced supercurrent on junction length in
  multilayer-graphene josephson junctions.
\newblock {\em Physica C: Superconductivity}, 470(20):1477--1480, 2010.

\bibitem{aliaga10}
Núria Aliaga-Alcalde, Patricia Marqu{\'e}s-Gallego, Mirte Kraaijkamp, Coral
  Herranz-Lancho, Hans den Dulk, Helmut Gorner, Olivier Roubeau, Simon~J Teat,
  Thomas Weyhermüller, and Jan Reedijk.
\newblock Copper curcuminoids containing anthracene groups: fluorescent
  molecules with cytotoxic activity.
\newblock {\em Inorganic chemistry}, 49(20):9655--9663, 2010.

\end{thebibliography}
\end{document}